# Electronic measurement and control of spin transport in Silicon

Ian Appelbaum[a], Biqin Huang[a], and Douwe Monsma[b]

[a] *Department of Electrical and Computer Engineering, University of Delaware, Newark, DE 19716*

[b] *Cambridge NanoTech, Inc., Cambridge, MA 02139*

**The electron spin lifetime and diffusion length are transport parameters that define the scale of coherence in spintronic devices and circuits. Since these parameters are many orders of magnitude larger in semiconductors[1,2] than in metals,[3,4] semiconductors could be the most suitable for spintronics. Thus far, spin transport has only been measured in direct-bandgap semiconductors[5] or in combination with magnetic semiconductors, excluding a wide range of non-magnetic semiconductors with indirect bandgaps. Most notable in this group is silicon (Si), which (in addition to its market entrenchment in electronics) has long been predicted a superior semiconductor for spintronics with enhanced lifetime and diffusion length due to low spin-orbit scattering and lattice inversion symmetry.[6-8] Despite its exciting promise, a demonstration of coherent spin transport in Si has remained elusive, because most experiments focused on magnetoresistive devices; these methods fail because of universal impedance mismatch obstacles,[9] and are obscured by Lorentz magnetoresistance and Hall effects.[10]**

**Here we demonstrate conduction band spin transport across 10 μm undoped Si, by using spin-dependent ballistic hot-electron filtering through ferromagnetic thin films for both spin-injection and detection. Not based on magnetoresistance, the hot electron spin-injection and detection avoids impedance mismatch issues and prevents interference from parasitic effects. The clean collector current thus shows**

**independent magnetic and electrical control of spin precession and confirms spin coherent drift in the conduction band of silicon.**

Figure 1(a) illustrates the operating principle and schematic band diagram of our device. Spin injection and detection is based on the attenuation of minority-spin hot electrons in ferromagnetic thin films, as in spin-valve transistors.[11,12] In our device, the spin-valve transistors used for injection and detection each only have a single ferromagnetic base layer and we define these as "hot-electron spin transistors". In step 1, a solid-state tunnel junction injects unpolarized electrons from the aluminum emitter into the ferromagnetic $Co_{84}Fe_{16}$ base, forming emitter current Ie. Spin-dependent hot-electron scattering attenuates minority spin electrons (step 2), so that the electrons transported over the Schottky barrier and into the undoped float-zone (FZ)-Si conduction band (forming injected current Ic1) are polarized, with their spin parallel to the magnetization of the $Co_{84}Fe_{16}$ (step 3).[13] After drift/diffusion through 10 μm of undoped Si (step 4), the spin polarization of the conduction band electrons is detected by a second hot-electron spin transistor. The $Ni_{80}Fe_{20}$ base again uses ballistic hot-electron spin filtering, so the "second collector current" (Ic2, step 5) formed from ballistic transport through the $Ni_{80}Fe_{20}$ and into the n-Si substrate conduction band, is dependent on the relative magnetizations of both ferromagnetic layers. When they are parallel, Ic2 is higher than when they are antiparallel, but only if electron spin polarization is maintained through the undoped Si layer. Therefore, this device is the electron analogue of the photon polarization-analyzer experiment in optics.

There are various *intrinsic* device aspects that allow a clean spin transport signal in Ic2, and that make it immune to fringe field-induced magnetoresistance and Hall effects: 1. The exponential spin selective mean free path dependence in the ferromagnetic films create very large spin polarizations. In principle this can approach 100%, allowing effective injection and detection at cryogenic and room temperatures[11];

2. Because the spin filtering is caused by bulk scattering in the ferromagnetic films, they are easy to reproduce, since there is no interface sensitivity to the spin filtering (as there is, e.g., in magnetic tunnel junctions); 3. This device, like a spin-valve transistor, is a high impedance current source.[11,12] Since Ic2 is driven by Ic1, and Ic1 by Ie, Ic2 is virtually independent of Vc1, the applied voltage across the Si drift region. This also means that any generated Hall voltage in the FZ-Si has no effect on Ic2. The underlying background to the insensitivity to resistance and voltage of the FZ-Si is that the potential is screened by the two Schottky barriers on either side, and that the electrons traveling in the FZ-Si conduction band are generated not by an ohmic source, but by hot-electron injection. 4. These devices operate over a wide temperature range, without appreciable change in Ic2, despite the fact that the resistivity of undoped Si varies by many orders of magnitude. The insensitivity of Ic2 to FZ-Si resistance implies that Ic2 is also insensitive to magnetoresistance in the FZ-Si. In fact, at the temperature we use for measurements here (85K), the FZ-Si is completely frozen out and its resistivity is $>10^{14}$ Ωcm. This means that there are no thermally or impurity generated electrons, and the only free electrons present are the injected spin polarized electrons!

There are several *design* aspects that provide our device with a clean spin-transfer current Ic2: 1. An undoped FZ-Si device layer is chosen because its extremely low impurity density results in wide Schottky depletion regions and a linear conduction band. This prevents potential wells and long spin dwell times; 2. The device is measured at 85K to eliminate Schottky leakage currents; 3. Copper (Cu) is used below the $Ni_{80}Fe_{20}$ to provide a low barrier and enable electrons injected by the higher Si-$Ni_{80}Fe_{20}$ Schottky barrier to overcome the Cu/n-Si Schottky barrier; 4. Shape anisotropy of the ferromagnetic films allows us to apply a perpendicular magnetic field for spin precession measurements, without orienting the magnetizations out of plane.

Our fabrication procedures described in the Methods section at the end of this paper are similar to those of the Spin-Valve Transisitor,[12] and result in an array of devices like the one wirebonded and shown in the micrograph displayed in Figure 1(b).

Figure 2 (a) shows the injector tunnel junction current-voltage characteristics, illustrating the expected non-linear Ie-Ve relationship. Figure 2 (b) shows the simultaneous measurement of injected current Ic1-Ve, demonstrating a threshold in the first collector current Ic1 at Ve=-0.8V. This represents the 0.8eV potential energy needed for the electrons to exceed the $Co_{84}Fe_{16}$/FZ-Si Schottky barrier height, and is typical for such metal base transistor-type structures. After vertical transport through 10 μm FZ-Si, some of these electrons travel ballistically through the $Ni_{80}Fe_{20}$/Cu film and into the n-Si collector, resulting in a second collector current (Ic2) detected at the In contact to the substrate. This signal rises above our detection limit at an emitter voltage of approximately -1.2V as shown in Figure 2(c).

In Figure 3(a), in-plane magnetic hysteresis data of Ic2 at Ve=-1.8V and 85K are shown. Measurements begin with fully saturated and aligned magnetizations by ramping to our magnetic field maximum. When the magnetic field is swept through zero and changes sign, first the $Ni_{80}Fe_{20}$ switches to align with the field and the magnetizations are anti-parallel, resulting in a reduction in Ic2 of approximately 2%, as shown in Fig. 3 (a). As the magnetic field passes through the $Co_{84}Fe_{16}$ switching field, the magnetizations again align and the higher collector current is regained by approximately 125 Oe, consistent with the magnetic hysteresis data shown in Fig. 3(b). The symmetric magnetic-field dependence of Ic2 upon reversal of sweep direction in Fig. 3(a) indicates that the electron spin maintains polarization while travelling through 10μm of FZ-Si. Results similar to Fig. 3 (a) are also found when the emitter tunnel junction is operated in constant current mode, and many devices have consistently been measured with substantially similar results.

In Figure 4 (a), we show the dependence of Ic2 on magnetic field ***perpendicular*** to the film plane with Vc1=0V. The measurement begins with the field at -5000Oe, where a small in-plane component is sufficient to saturate the in-plane $Ni_{80}Fe_{20}$ and $Co_{84}Fe_{16}$ magnetizations in a parallel configuration. Following the red line to the right towards smaller field values, we notice a peak near -700Oe and a dip near -350Oe. Once the magnetic field increases toward positive values past zero, the small in-plane component of external magnetic field switches the $Ni_{80}Fe_{20}$ magnetization, causing an antiparallel magnetization configuration and inverting the magnitudes of the oscillation. Reversing the field scan direction (blue line) results in similar features reflected on the magnetic field axis, but with a deeper dip at small negative field due to sharper $Ni_{80}Fe_{20}$ switching in that direction.

Although Vc1=0V in Fig 4(a), a vertical drift field (due to a voltage drop caused by emitter current flowing through the Al/$Co_{84}Fe_{16}$ thin base film) exists in the FZ-Si. Under these conditions, where drift is the primary transport mechanism, the magnetic fields of the extrema seen in Fig. 4(a) can be identified approximately as the conditions for $\pi$ and $2\pi$ precession. The precessing spin direction, projected on the measurement axis defined by the $Ni_{80}Fe_{20}$ magnetization, causes oscillation in Ic2.

Because of the action of random diffusion, spin dephasing occurs with higher precession angles, and results in dampened higher-order oscillations. Therefore, just two multiples of $\pi$ precession angle are seen. A useful advantage of the four-terminal design and rectifying Schottky barriers on either side of the FZ-Si in our device is that Ic2 is virtually independent of voltage across the drift region. Therefore, the drift electric field can be tuned to change the spin polarized electron transit time between injection and detection with applied voltage Vc1. When the electric drift field is increased with an applied voltage bias, the transit time is reduced, and the precession angle at any fixed magnetic field is consequently also reduced. This pushes the extrema to higher values of perpendicular magnetic fields, clearly shown in Fig. 4(b) and (c) at 0.5V and 1.0V accelerating voltage. Under these conditions of higher accelerating electric field, drift is even more dominant than in Fig. 4(a), and precession angles up to 4π can be seen. The precession extrema for parallel magnetization configurations are labelled with their precession angle in the Figs. 4(a-c). In this way, maxima correspond to even multiples of π and minima correspond to odd multiples of π. *The large number of precession extrema and their expected electric field dependence are conclusive evidence of strong coherent spin transport in the FZ-Si.*

Using the spin precession frequency $g\frac{\mu_B}{h} = 2.8\text{MHz/Oe}$,[14] the transit time can be calculated from the magnetic field value of the first extrema, where the average precession angle is approximately π radians. Using ±350 Oe as the position of the extrema in Fig. 4(a), we have $\tau_{transit} = \frac{\pi}{2\pi \cdot 2.8\times10^6 * 350} \approx 0.5ns$, which gives an approximately $2\times10^6$ cm/s drift velocity through the 10 μm thick FZ-Si.[15] The positions of the precession extrema in Fig. 4 (b) and (c) indicate transit times of ~0.3 ns and ~0.2 ns, respectively.

Since the electron transit time is controlled by applied voltage Vc1, the normalized magnetocurrent change in Ic2 (ΔIc2) can be used to deduce the spin lifetime in FZ-Si. Assuming a simple exponential decay law, and using the measured transit times from Fig. 4(a-c), we can calculate the spin lifetime $\tau_{sf}$ (at 85K) using ratios of

$$\frac{\Delta I_{c2}(V_{c1})}{I_{c1}(V_{c1})} \propto e^{-\frac{\tau_{transit}(V_{c1})}{\tau_{sf}}},$$

for each pair of data. With the maximum fluctuations shown in Fig. 4 (a-c), and using simultaneously-measured Ic1(0V)=118nA, Ic1(0.5V)=132nA, and Ic1(1.0V)=139nA, we calculate $\tau_{sf} \approx 1ns$. This value is consistent with spin lifetimes in the direct-bandgap semiconductor GaAs measured at similar temperatures using optical techniques.[1] However, we must consider this value a lower bound, with the actual lifetime potentially much longer: the applied voltage Vc1 could alter the magnetocurrent magnitude separately from the finite spin lifetime and variable transit time effect, reducing our estimate.

It is possible to employ alternative spin injection schemes with our detection method, such as direct tunnel injection[16,17,18]. However, in this case the injection current is not decoupled from Vc1 and thus independent control of spin precession by magnetic field and electric drift field is compromised; on the other hand, the total signal will be larger. Finally, we hope that this injection/detection scheme will be implemented in lateral geometries, to advance the development of future spin based integrated circuits.

**Methods:**

We fabricated this device using room-temperature ultra-high-vacuum wafer bonding[12] (as in regular spin-valve transistors) to assemble the semiconductor-metal-semiconductor spin detection structure. First, a 4 nm-thick Cu film is deposited via thermal evaporation on a 1 Ω·cm 50mm diameter n-Si wafer to achieve a low second collector Schottky barrier. Then, during deposition of 2nm of $Ni_{80}Fe_{20}$ on both this wafer and a 50mm diameter silicon-on-insulator (SOI) wafer with 10 μm single-crystal FZ-Si device layer, the wafers are pressed together *in-situ*, forming a cohesive bond and a single 4nm-thick $Ni_{80}Fe_{20}$ film. Transmission electron microscopy (TEM) analysis of similarly bonded interfaces is shown in Ref. 12. After handle substrate removal with Tetra-Methyl Ammonium Hydroxide and mesa patterning using the 2 μm-thick buried oxide as a mask, a tunnel junction structure is deposited using electron beam deposition of a 5nm $Co_{84}Fe_{16}$ and 5nm Al bilayer 500x500 $\mu m^2$ base and (after UV ozone treatment to form the tunnel barrier) two Al contacts. For isolation, a wafer saw is used to cut trenches partially through the n-Si collector substrate, forming an array of devices with 900x900 $\mu m^2$ lateral size.

Pinholes in these deposited layers cannot result in spurious measurements of Ic2: 1. If they exist in the tunnel oxide, the emitter tunnel junction is shorted and the device is clearly inoperable; 2. If they exist in the Al/$Co_{84}Fe_{16}$ layer, the emitter forms a rectifying Schottky barrier there and the device is unaffected; and 3. Pinholes in the single-crystal FZ-Si layer are not possible.

As fabricated, the two Al contacts both form tunnel junctions with the Al/ $Co_{84}Fe_{16}$ base. Since tunnel junctions break down and electrically short at high applied voltages, ramping a current source through them in series breaks one and leaves the other intact. Contact for Ic1 is to the $Ni_{20}Fe_{80}$ buried layer, and contact for Ic2 is with cold-pressed indium to the n-Si substrate. This current has been measured with a standard, commercially available, Keithley 236 source-measure unit, without any other additions or special efforts.

**Acknowledgements:**

We acknowledge assistance during fabrication from Igor Altfeder, SQUID measurements by George Hadjipanayis and Alexander Gabay, and use of the wafer saw from Keith Goossen. This work is supported by ONR and DARPA/MTO.

Correspondence and requests for materials should be addressed to I.A. (appelbaum@ee.udel.edu).

## Figure Legends:

Figure 1 (a) and (b): Illustration of the Si spin transport device. (a) Schematic band diagram of the Si electron spin transport device. We operate the device in constant emitter voltage (Ve), measuring the “first collector current” (Ic1) at the NiFe contact and “second collector current” (Ic2) at an In contact to the n-Si substrate, under optional voltage bias (Vc1) across the single-crystal float-zone Si (FZ-Si) drift region. Refer to text for explanation of sequential transport steps (1)-(5). (b) A top-down micrograph of a representative wire-bonded Si spin-transport device, showing the device structure, contacts to the spin-injection tunnel junction (TJ) base and emitter, and spin-detector buried NiFe layer.

Figure 2 (a)-(c): Simultaneously measured current dependence on tunnel-junction emitter voltage at 85K. (a) emitter current; (b) first collector current Ic1 at the $Ni_{80}Fe_{20}$ contact for Vc1=0V; (c) second collector current Ic2 at an In contact to n-Si substrate.

Figure 3 (a) and (b): In-plane magnetic field dependence at 85K. Arrows indicate magnetic field sweep direction. (a) Second collector current Ic2 at constant emitter bias $V_e$=-1.8V and Vc1=0V, showing a clear ~2% spin-valve effect; and (b) SQUID magnetometer measurements, showing switching fields consistent with the behaviour seen in (a).

Figure 4 (a)-(c): Spin precession and dephasing in a perpendicular magnetic field at constant emitter voltage $V_e$=-1.8V and 85K. (a) second collector current Ic2 in zero applied voltage Vc1; (b) under accelerating voltage bias of 0.5V; and (c) under accelerating voltage bias of 1.0V. At higher voltages, the spin polarized electron transport time is reduced in the increased drift field, so precession angle at fixed field is smaller, causing increased precession period and revealing presence of precession angles up to $4\pi$.

Appelbaum et al., Fig 1

(a)

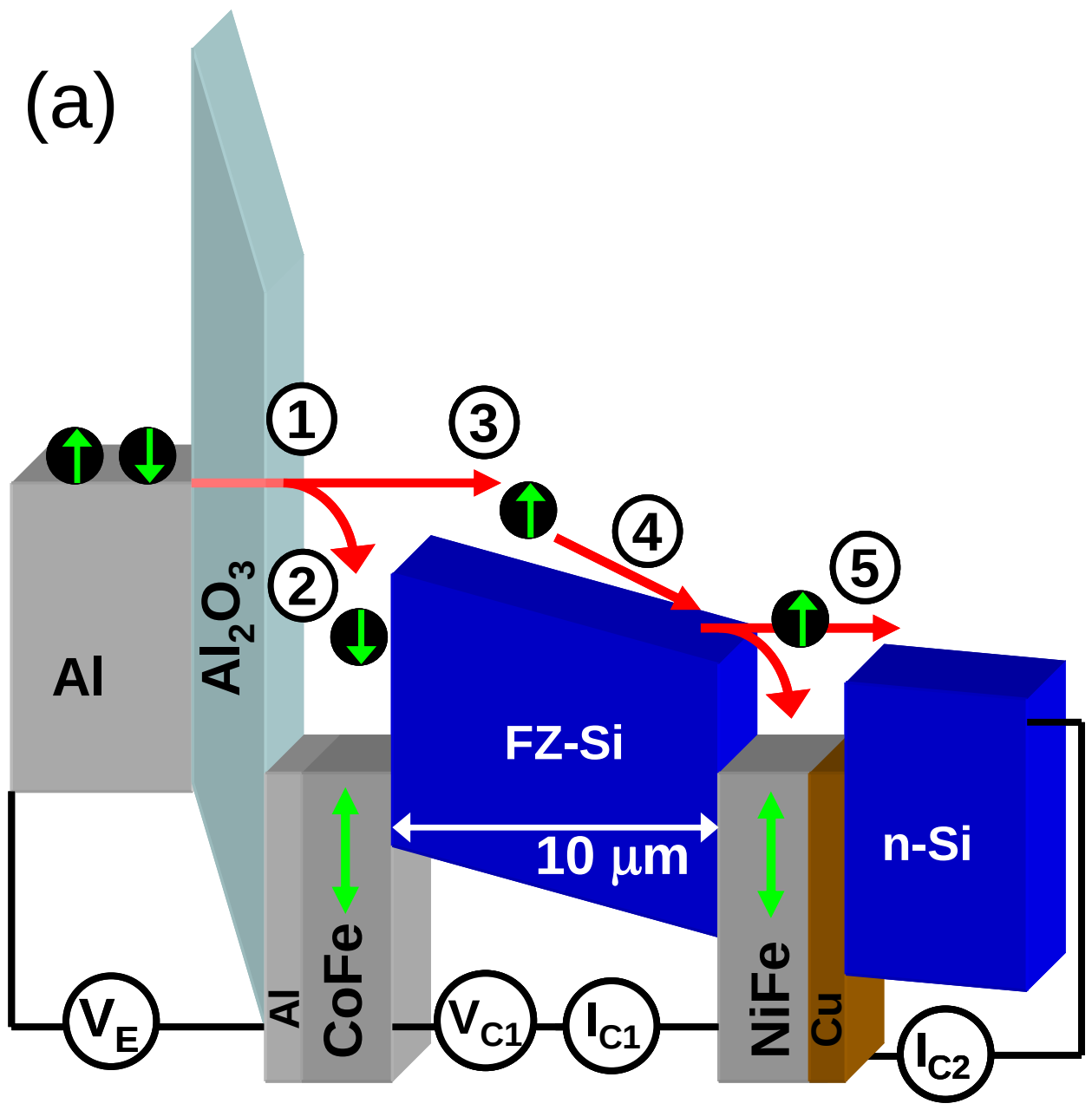

(b)

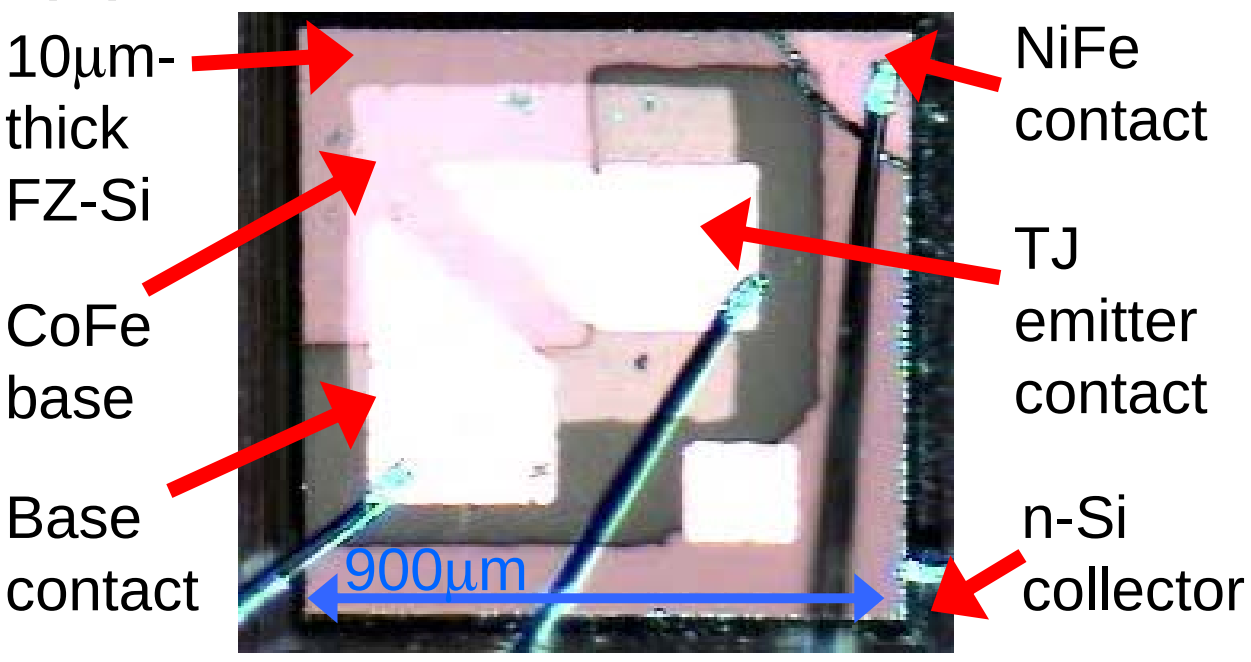

Appelbaum et al., Fig 2

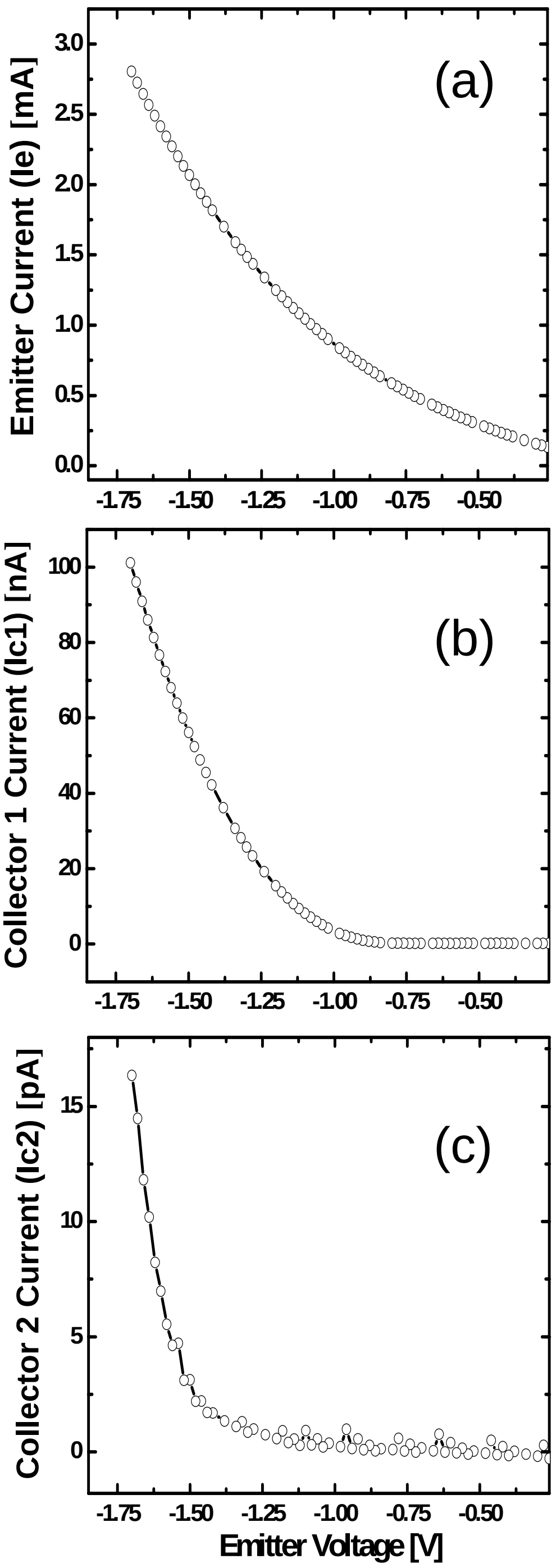

Appelbaum et al., Fig 3

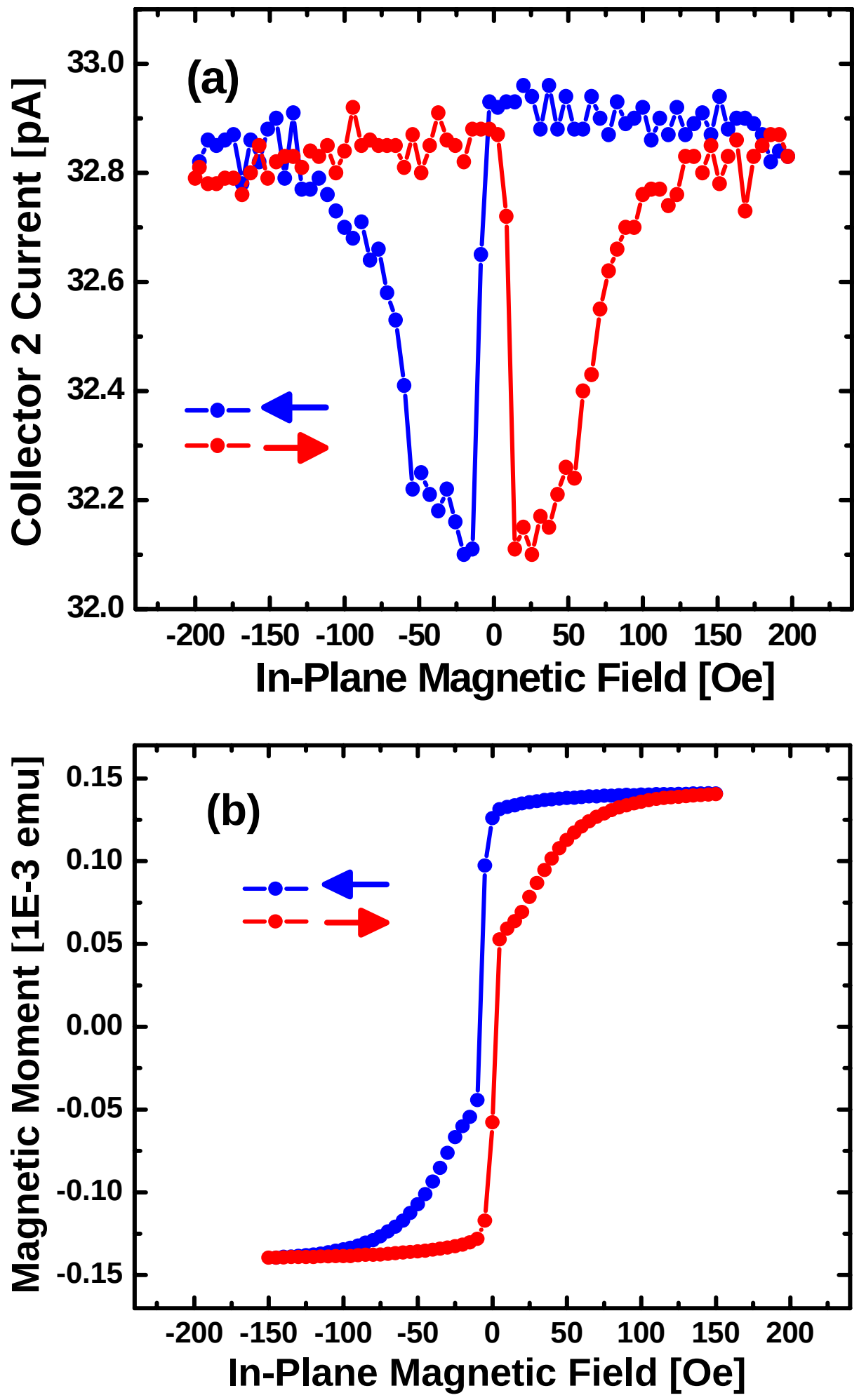

Appelbaum et al., Fig 4

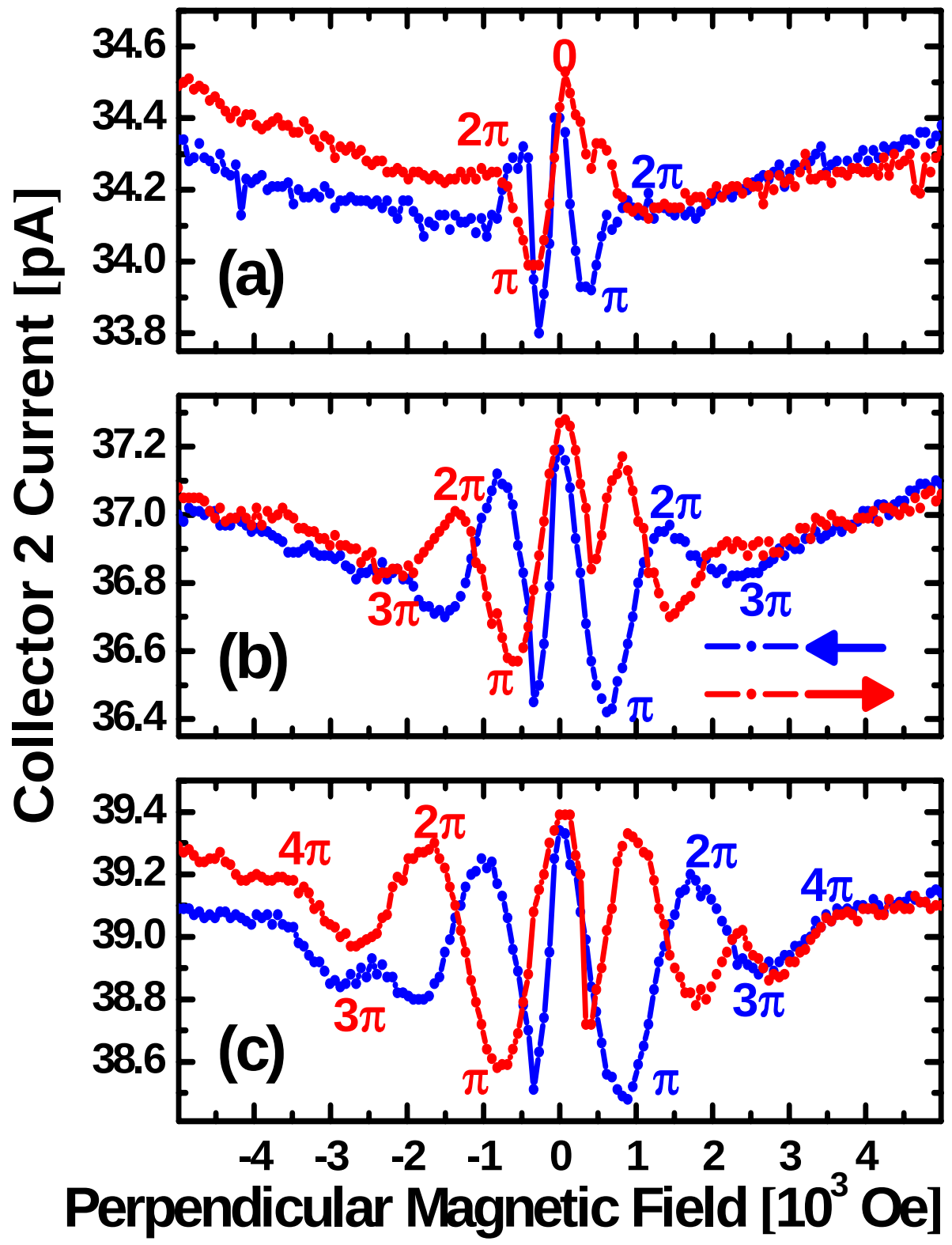